\documentclass{moriond}
\usepackage{amsmath}

\def\be{\begin{equation}}
\def\ee{\end{equation}}
\def\bea{\begin{eqnarray}}
\def\eea{\end{eqnarray}}

\newcommand{\Photo}{}

\begin{document}
\rightline{DO-TH 21/14}
\vspace*{4cm}
\title{PHYSICS REACH OF ${D}_{(s)}\to \pi(K) \ell\ell$ AND OTHER CHARMING NULL TEST OPPORTUNITIES}

\author{{MARCEL GOLZ$^1$}}

\address{$^1$Fakultät für Physik, TU Dortmund, Otto-Hahn-Str.~4, D-44221 Dortmund, Germany}

\maketitle\abstracts{
We discuss possibilities to test physics beyond the Standard Model in $\vert\Delta c\vert=\vert\Delta u\vert= 1$ semileptonic, hadronic and missing energy decay modes. Clean null test observables such as angular observables, CP-asymmetries and lepton universality tests are presented and model-independent correlations as well as details within flavorful, anomaly-free $Z^\prime$ models are worked out. }

\section{Introduction}

Flavor Changing Neutral Current (FCNC) charm decays are sensitive probes of physics beyond the Standard Model (BSM). The underlying $c\to u \ell\ell$ quark transition is the up-type cousin of the down-type FCNC transitions $b\to s\ell\ell$ and $s\to d \ell \ell$. With down-type quarks in the leading loop diagrams, the up-type sector suffers from a severe GIM cancellation. Together with the loop suppression, the GIM mechanism leads to tiny Standard Model (SM) contributions to rare charm decays and in turn to a high sensitivity to New Physics (NP).

In the SM, semileptonic decays, such as $D^+\to\pi^+\ell^+ \ell^-$, $D_s^+\to K^+\ell^+\ell^-$, are dominated by long-range dynamics with sizable hadronic uncertainties. Therefore, NP searches in branching ratios are challenging as the NP nature and long-range QCD parameters need to be disentangled simultaneously.

\begin{figure}[h!]
\begin{minipage}{0.99\linewidth}
\centering{\includegraphics[width=0.37\linewidth]{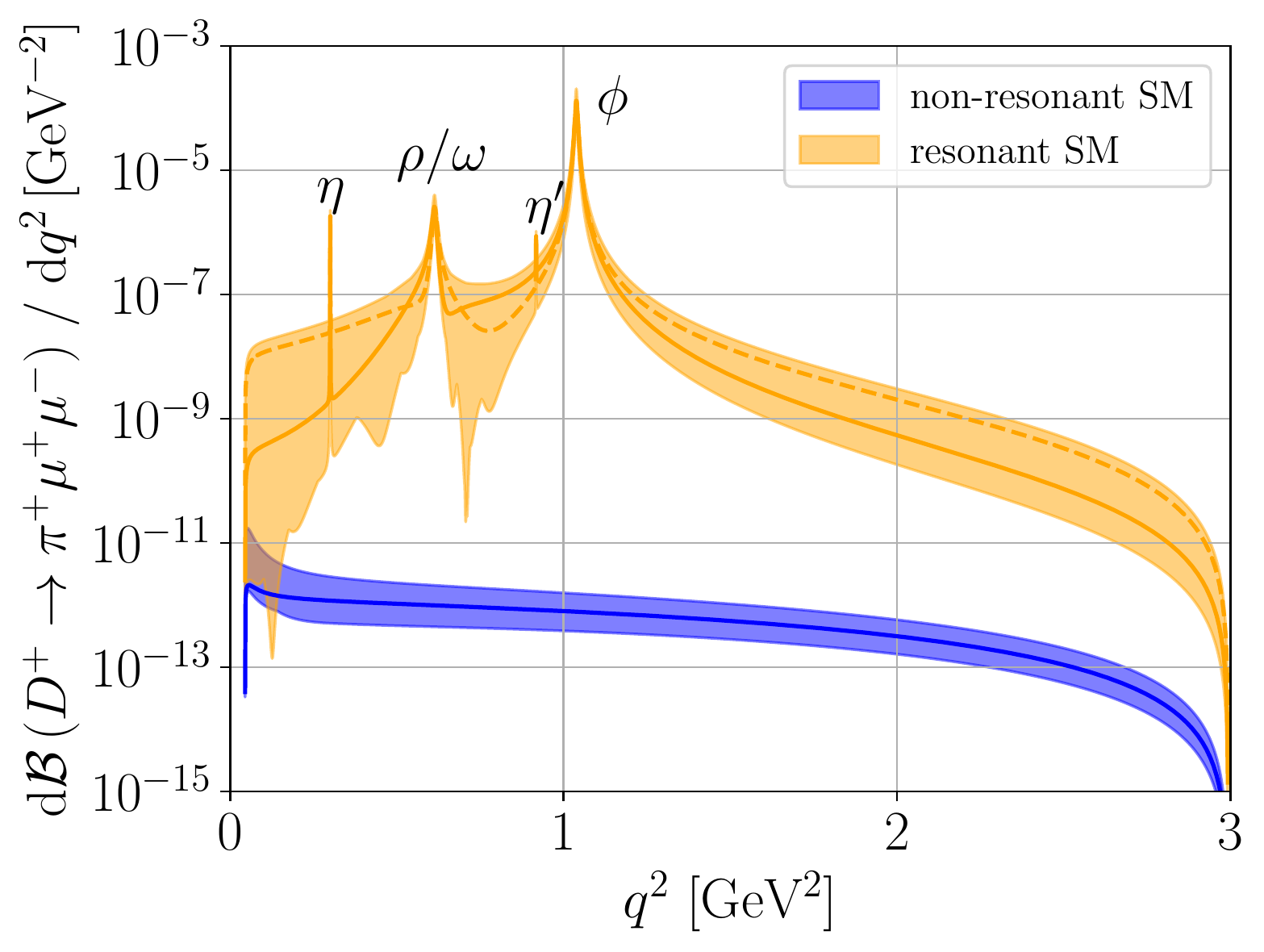}}
\end{minipage}
\hfill
\caption[]{SM differential branching ratio for $D^+\to\pi^+\mu^+\mu^-$. (Non-)resonant contributions are shown in orange (blue). Uncertainties arise in the resonant contributions from the model of a sum of Breit-Wigner shapes including unknown strong phases and from scale uncertainties of the SM Wilson coefficients in the non-resonant case. Both contributions include uncertainties from form factors. Figure taken from Ref.~\cite{Bause:2019vpr}.}
\label{fig:smbr}
\end{figure}

As a prime example, the dilepton invariant mass ($q^2$) differential branching ratio of the $D^+\to\pi^+\mu^+\mu^-$ decay is shown in Fig.~\ref{fig:smbr}. Clearly, contributions from intermediate $\rho,\, \omega,\, \eta,\, \phi$ and $\eta^\prime$ resonances (indicated in orange) dominate the perturbative SM contributions (in blue) in the full kinematic range. Therefore, a non-resonant region does not exist in rare charm decays. Measurements of the resonance tails are still useful, as they provide information on the underlying QCD dynamics that otherwise produce huge uncertainties, as evident from the width of the orange bands.

On the other hand, the GIM mechanism also leads to $C_{10}=0$, such that any observable proportional to $C_{10}$ constitutes a clean null test of the SM. 

Recent experimental and theoretical progress is summarized in Ref.~\cite{Gisbert:2020vjx}. This includes upper limits such as $\mathcal{B}(D^+\to\pi^+\mu^+\mu^-)<6.7\times 10^{-8}$ at $90\,\%$ CL~\cite{Aaij:2020wyk} and $\mathcal{B}(\Lambda_c\to p \ell\ell)<7.7\times 10^{-8}$ at $90\,\%$ CL~\cite{Aaij:2017nsd}, first measurements in $D^0\to P_1 P_2 \ell\ell$~\cite{Aaij:2017iyr} and the first observation of CP-violation in the charm sector in $\Delta A_{{\rm CP}}$~\cite{Aaij:2019kcg}. On the theoretical side research interest is increasing and ongoing, see Refs.~\cite{Fajfer:2012nr,deBoer:2015boa,Fajfer:2015mia,Feldmann:2017izn,Meinel:2017ggx,deBoer:2018buv,Adolph:2018hde,Bause:2019vpr,Bause:2020obd,Adolph:2020ema,Bause:2020auq,Bause:2020xzj,Bharucha:2020eup,Adolph:2021ncg}.

The aim of this letter is to present null test strategies that overcome the issue of large uncertainties from non-perturbative origin and therefore provide clean tests of the SM.

\section{Null test strategies in rare charm decays}

\subsection{Angular observables}
\begin{figure}[h!]
\begin{minipage}{0.49\linewidth}
\centerline{\includegraphics[width=0.74\linewidth]{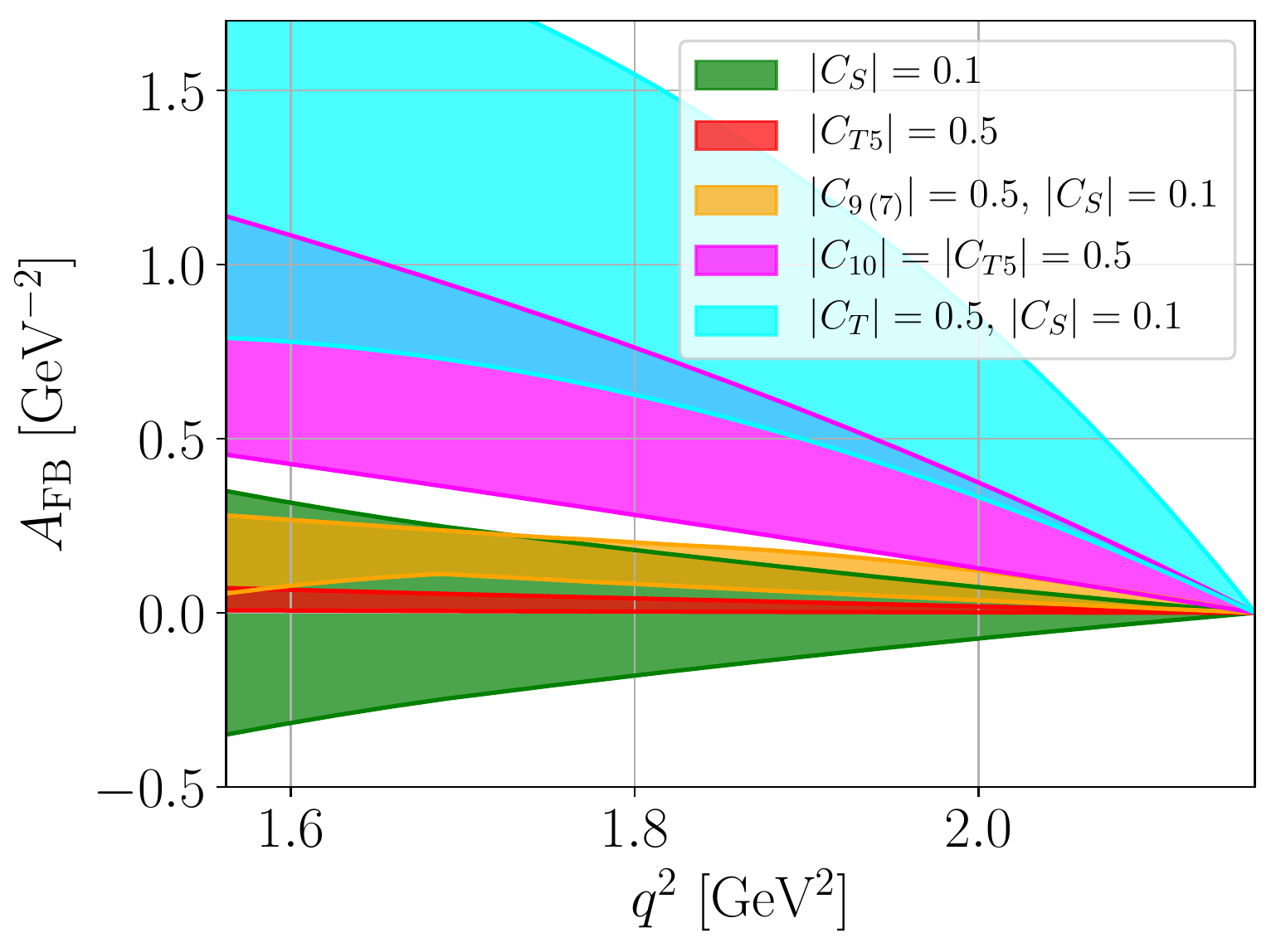}}
\end{minipage}
\hfill
\begin{minipage}{0.49\linewidth}
\centerline{\includegraphics[width=0.7\linewidth]{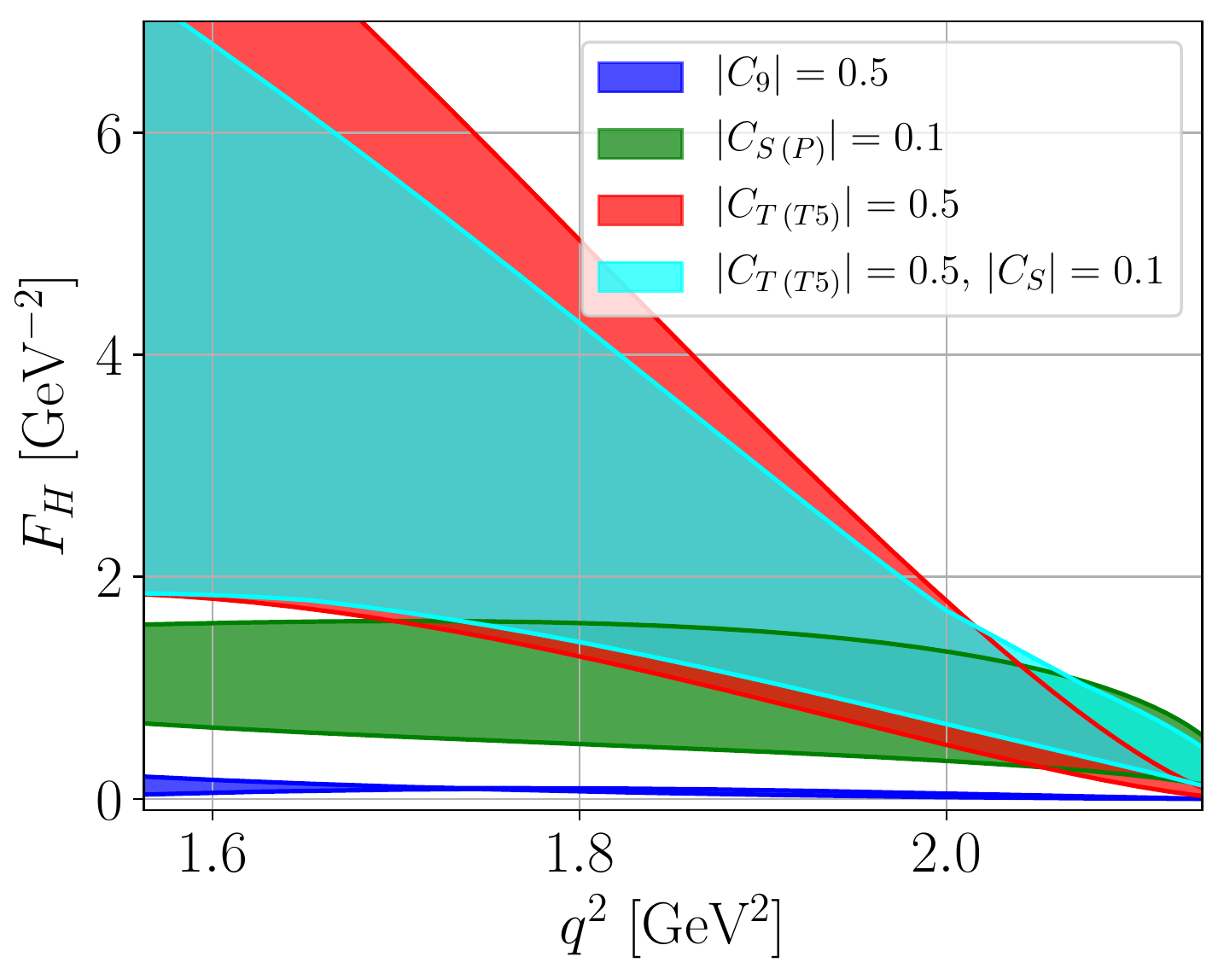}}
\end{minipage}
\hfill
\caption[]{Lepton forward backward asymmetry $A_{\rm FB}$ (left) and flat term $F_H$ (right) for various benchmark contribution to Wilson coefficients. The SM is zero in the left plot and sufficiently below the blue curve in the right plot and therefore not shown. Figures taken from Ref.~\cite{Bause:2019vpr}.}
\label{fig:angl}
\end{figure}

The first class of observables is constructed from the full angular distribution of $D\to P \ell\ell$
\begin{equation}
\frac{1}{\Gamma_{\ell}}\,\frac{{\rm d}\Gamma_\ell}{{\rm d}\cos\theta}=\frac{3}{4}(1-F_H)(1-\cos^2\theta_\ell) + \frac{1}{2}F_H + A_{\rm FB}\cos \theta_\ell\,,
\end{equation}
where $\Gamma_\ell=\int_{q^2_{\rm min}}^{q^2_{\rm max}}{\rm d}\Gamma / {\rm d}q^2 \,\text{d}q^2$ and $\theta_l$ denotes the angle of the $\ell^-$ and the $D$ meson in the dilepton rest frame. 
These two angular observables are the lepton forward backward asymmetry, $A_{\rm FB}$, and the flat term, $F_H$, which are shown in Fig.~\ref{fig:angl} in the high $q^2$ region (above the $\phi$ resonance) left and right plot, respectively.
$A_{\rm FB}$ vanishes in the SM and only receives contributions from interference terms between two different Wilson coefficients (WCs). Hence, any signal is a clear sign of BSM physics. Largest contributions are obtained for combinations involving scalar and tensor WCs, as they reveive no suppression by the light lepton mass $m_\ell$. The flat term $F_H$ is not zero in the SM, however only contributing via $m_\ell$ suppressed terms, which are small in the muon case and negligible for electrons in the final state. Similar to $A_{\rm FB}$, tensor and scalar contributions do not suffer from this suppression and produce signal sufficiently above the SM expectations.

\subsection{CP asymmetries}
\begin{figure}[h!]
\begin{minipage}{0.49\linewidth}
\centerline{\includegraphics[width=0.74\linewidth]{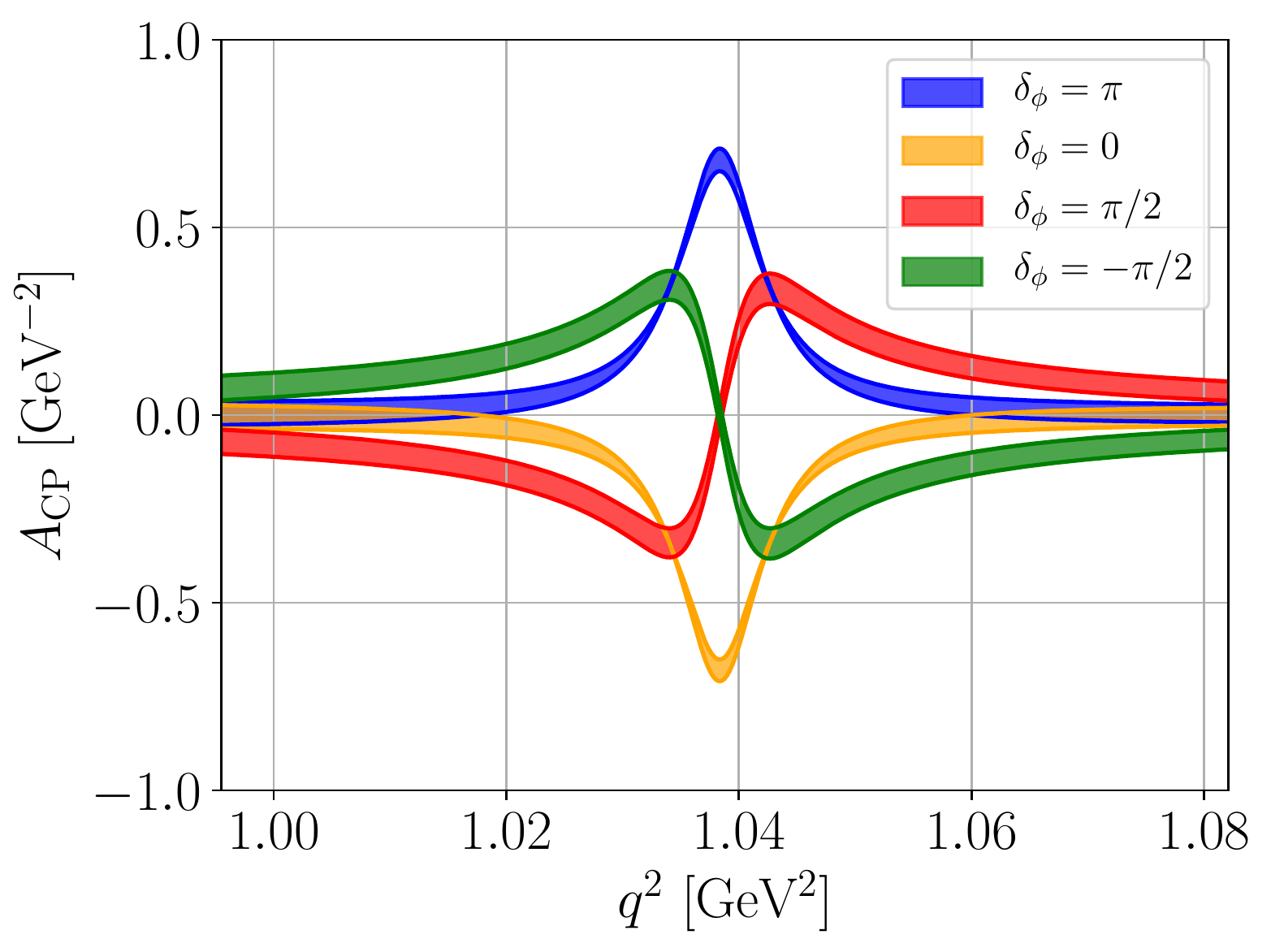}}
\end{minipage}
\hfill
\begin{minipage}{0.49\linewidth}
\centerline{\includegraphics[width=0.74\linewidth]{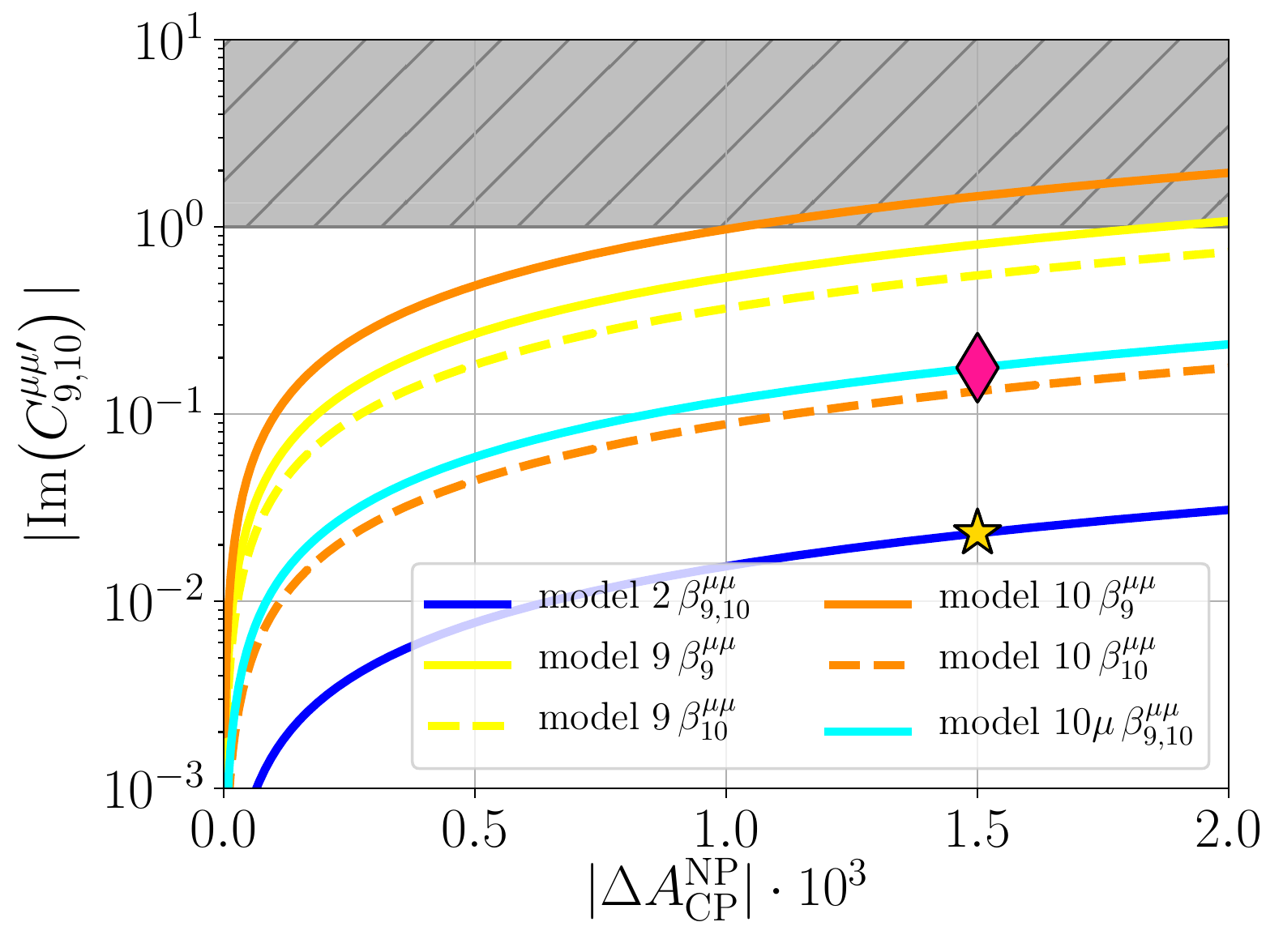}}
\end{minipage}
\hfill
\caption[]{The left plot shows the CP asymmetry around the $\phi$ resonance for $D_s^+\to K^+ \mu^+\mu^-$ and a benchmark value of $C_9=0.1\,\exp(i\frac{\pi}{4})$ and various different fixed strong phases taken from Ref.~\cite{Bause:2019vpr}. The right plot displays the correlation between large values of $\Delta A_{\rm CP}$ and Im$(C_{9,\,10})$ within different flavorful, anomaly-free $Z^\prime$ models. The right plot is taken from Ref.~\cite{Bause:2020obd}.}
\label{fig:acp}
\end{figure}

CP asymmetries are sizable at the resonances due to interference of possible CP violating BSM contributions with the resonance contribution. First pointed out in Ref.~\cite{Fajfer:2012nr}, this resonance catalyzed CP asymmetry has been promoted as a null test in the literature many times~\cite{deBoer:2015boa,Fajfer:2015mia,deBoer:2018buv,Bause:2019vpr}, as the SM contribution is negligible due to small CKM phases and, again, the strong GIM suppression. A benchmark scenario around the $\phi$ resonance is shown in the left plot of Fig.~\ref{fig:acp} for $D_s^+\to K^+ \mu^+\mu^-$. Recently, the LHCb collaboration has reported the first observation of charm CP violation in hadronic decays~\cite{Aaij:2019kcg}, where the NP nature of the measurement is not settled. In Ref.~\cite{Bause:2020obd} the connection between these two CP violating observables is studied in the framework of flavorful, anomaly-free $Z^\prime$ models and a strong correlation is found. As evident from the right plot in Fig.~\ref{fig:acp}, a large NP contribution to $\Delta A_{\rm CP}$ coincides with large CP violating effects in semileptonic decays. The benchmark in the left plot of Fig.~\ref{fig:acp} is $\vert{\rm Im}(C_9)\vert\sim 0.1$.
\subsection{Lepton universality tests}
Lepton universality (LU) is deeply rooted in the SM, however put to test in $b\to s$ transitions with increasing significance~\cite{Aaij:2021vac}. Similar tests of LU ratios are also possible in rare charm decays, where sizable effects are possible due to enhancements originating from the interference of NP with the resonant contributions~\cite{Fajfer:2015mia,Bause:2019vpr,deBoer:2018buv}.

A new approach to test LU was presented recently in Refs.~\cite{Bause:2020auq,Bause:2020xzj}. Here, dineutrino modes, which in the case of rare charm decays are null tests of the SM themselves and do not suffer from resonance domination, are constrained utilizing different flavor assumptions in the charged lepton sector and exploiting the ${\rm SU}(2)_L$ link between charged leptons and neutrinos in the Standard Model Effective Field Theory (SMEFT). For example upper limits for $\mathcal{B}(D^+\to\pi^+\bar{\nu}\nu)$ are obtained assuming LU, charged lepton flavor conservation (cLFC) and without assumptions (general):
\begin{equation}
\begin{split}
\mathcal{B}(D^+\to\pi^+\bar{\nu}\nu) &< 2.5\times10^{-6} \quad {\rm(LU)}\,,\\
\mathcal{B}(D^+\to\pi^+\bar{\nu}\nu) &< 1.4\times10^{-5}  \quad {\rm(cLFC)}\,,\\
\mathcal{B}(D^+\to\pi^+\bar{\nu}\nu) &< 5.2\times10^{-5}  \quad {\rm(general)}\,.
\end{split}
\end{equation}
A measurement above one of the respective bounds implies the breakdown of the assumed flavor symmetry.

\section{Conclusion}

We have presented null test strategies for the search for NP with rare charm decays. Huge uncertainties of resonant contributions dominating the decay modes of interest can be overcome in angular observables, CP-asymmetries, lepton universality ratios and dineutrino modes. Synergies, such as the correlation between CP violating observables in hadronic and semileptonic decays or the ${\rm SU}(2)_L$ link between charged leptons and neutrinos were presented and provide a formidable road to observe NP in rare charm decays. 

\section*{Acknowledgments}
Many thanks to the organizers for the opportunity to present this work and for coordinating this special format of Moriond@home. I am grateful to  Gudrun Hiller, Rigo Bause and Hector Gisbert for fruitful collaboration and for reading this manuscript.
MG is supported by the "Studienstiftung des deutschen Volkes".


\section*{References}

\begin{thebibliography}{99}
\bibitem{Gisbert:2020vjx}
H.~Gisbert, M.~Golz and D.~S.~Mitzel,
Mod. Phys. Lett. A \textbf{36} (2021) no.04, 2130002
doi:10.1142/S0217732321300020
[arXiv:2011.09478 [hep-ph]].

\bibitem{Aaij:2020wyk}
R.~Aaij \textit{et al.} [LHCb],
[arXiv:2011.00217 [hep-ex]].


\bibitem{Aaij:2017nsd}
R.~Aaij \textit{et al.} [LHCb],
Phys. Rev. D \textbf{97} (2018) no.9, 091101
doi:10.1103/PhysRevD.97.091101
[arXiv:1712.07938 [hep-ex]].


\bibitem{Aaij:2017iyr}
R.~Aaij \textit{et al.} [LHCb],
Phys. Rev. Lett. \textbf{119} (2017) no.18, 181805
doi:10.1103/PhysRevLett.119.181805
[arXiv:1707.08377 [hep-ex]].


\bibitem{Aaij:2019kcg}
R.~Aaij \textit{et al.} [LHCb],
Phys. Rev. Lett. \textbf{122} (2019) no.21, 211803
doi:10.1103/PhysRevLett.122.211803
[arXiv:1903.08726 [hep-ex]].



\bibitem{Fajfer:2012nr}
S.~Fajfer and N.~Ko\v{s}nik,
Phys. Rev. D \textbf{87} (2013) no.5, 054026
doi:10.1103/PhysRevD.87.054026
[arXiv:1208.0759 [hep-ph]].


\bibitem{deBoer:2015boa}
S.~de Boer and G.~Hiller,
Phys. Rev. D \textbf{93} (2016) no.7, 074001
doi:10.1103/PhysRevD.93.074001
[arXiv:1510.00311 [hep-ph]].

\bibitem{Fajfer:2015mia}
S.~Fajfer and N.~Ko\v{s}nik,
Eur. Phys. J. C \textbf{75} (2015) no.12, 567
doi:10.1140/epjc/s10052-015-3801-2
[arXiv:1510.00965 [hep-ph]].

\bibitem{Feldmann:2017izn}
T.~Feldmann, B.~M\"uller and D.~Seidel,
JHEP \textbf{08} (2017), 105
doi:10.1007/JHEP08(2017)105
[arXiv:1705.05891 [hep-ph]].

\bibitem{Meinel:2017ggx}
S.~Meinel,
Phys. Rev. D \textbf{97} (2018) no.3, 034511
doi:10.1103/PhysRevD.97.034511
[arXiv:1712.05783 [hep-lat]].

\bibitem{deBoer:2018buv}
S.~De Boer and G.~Hiller,
Phys. Rev. D \textbf{98} (2018) no.3, 035041
doi:10.1103/PhysRevD.98.035041
[arXiv:1805.08516 [hep-ph]].



\bibitem{Adolph:2018hde}
N.~Adolph, G.~Hiller and A.~Tayduganov,
Phys. Rev. D \textbf{99} (2019) no.7, 075023
doi:10.1103/PhysRevD.99.075023
[arXiv:1812.04679 [hep-ph]].

\bibitem{Bause:2019vpr}
R.~Bause, M.~Golz, G.~Hiller and A.~Tayduganov,
Eur. Phys. J. C \textbf{80} (2020) no.1, 65
[erratum: Eur. Phys. J. C \textbf{81} (2021) no.3, 219]
doi:10.1140/epjc/s10052-020-7621-7
[arXiv:1909.11108 [hep-ph]].



\bibitem{Bause:2020obd}
R.~Bause, H.~Gisbert, M.~Golz and G.~Hiller,
Phys. Rev. D \textbf{101} (2020) no.11, 115006
doi:10.1103/PhysRevD.101.115006
[arXiv:2004.01206 [hep-ph]].




\bibitem{Adolph:2020ema}
N.~Adolph, J.~Brod and G.~Hiller,
Eur. Phys. J. C \textbf{81} (2021) no.1, 45
doi:10.1140/epjc/s10052-021-08832-3
[arXiv:2009.14212 [hep-ph]].


\bibitem{Bause:2020auq}
R.~Bause, H.~Gisbert, M.~Golz and G.~Hiller,
[arXiv:2007.05001 [hep-ph]].

\bibitem{Bause:2020xzj}
R.~Bause, H.~Gisbert, M.~Golz and G.~Hiller,
Phys. Rev. D \textbf{103} (2021) no.1, 015033
doi:10.1103/PhysRevD.103.015033
[arXiv:2010.02225 [hep-ph]].

\bibitem{Bharucha:2020eup}
A.~Bharucha, D.~Boito and C.~M\'eaux,
[arXiv:2011.12856 [hep-ph]].

\bibitem{Adolph:2021ncg}
N.~Adolph and G.~Hiller,
[arXiv:2104.08287 [hep-ph]].


\bibitem{Aaij:2021vac}
R.~Aaij \textit{et al.} [LHCb],
[arXiv:2103.11769 [hep-ex]].







\end{thebibliography}
\end{document}